\documentstyle[preprint,prb,aps]{revtex}
\begin{document}
\draft
\title
{Order-N Density-Matrix Electronic-Structure Method\\
for General Potentials}
\author{A. E. Carlsson}
\address{Department of Physics\\
Washington University\\
St. Louis, Missouri 63130-4899\\}
\date{\today}
\maketitle
\input epsf
\begin{abstract}
A new order-N method for calculating the electronic structure
of general (non-tight-binding) potentials is presented.
The method uses a combination of the ``purification''-based
approaches used by Li, Nunes and Vanderbilt, and Daw,
and a representation of the density matrix based on
``travelling basis orbitals''. This method gives a total energy
form that has the form of a cubic multicomponent Landau theory.
The method is applied to several one-dimensional examples,
including the free electron gas, the ``Morse'' bound-state
potential, a discontinuous potential that mimics an interface,
and an oscillatory potential that mimics a semiconductor.
The method is found to contain several physical effects that
are hard to obtain in real-space total-energy functionals:
Friedel oscillations, quantization of charge in bound states,
and band gap formation. Quantitatively accurate agreement with
exact results is found in most cases.  Possible advantages with
regard to treating electron-electron interactions and arbitrary
boundary conditions are discussed.
\end{abstract}
\vskip 0.5cm

\section{Introduction}

Recent years have seen the introduction of a number of fast
``order-N'' methods for calculating electronic properties,
total energies, and forces corresponding to complex atomic
configurations in materials.  The major motivation behind
these is to be able to perform molecular-dynamics type simulations
with forces that correctly reflect the electronic structure.
The methods have used a broad range of physical approaches.
The earliest ones involved local solution of the Schroedinger
equation in different regions of space,\cite{wyang91}
and discretization of the kinetic-energy operator combined with
subsequent recursion-based calculation of the electronic
Green's function.\cite{baroni92} Later methods include
transformation of the Kohn-Sham equations to a localized-orbital
representation,\cite{galli92} an iteratively obtained description
of the occupied subspace,\cite{stechel94} and two approaches
in which the electronic density matrix is explicitly solved
for in a sparse representation.\cite{xpLi93,daw93}

The present method builds on the last two references. The density
matrix is an operator that contains all of the information
about the electronic wave functions. For a review of the
density matrix in molecular systems, see Ref.~7.
One of the earliest applications of the density matrix to
condensed-matter systems was by Smith and Gay.\cite{smith75}
Because the density matrix decays (although not necessarily
very rapidly) as a function of separation, a truncation can be
used to obtain an order-N method.  Using a variational
principle\cite{aec83} for the density-matrix together with
a ``purification scheme''\cite{xpLi93} and a closely related
approach\cite{daw93} order-N methods for tight-binding models
have been developed.  The variational density-matrix method
has the advantage over recursion-type methods, which are also
order-N, that forces which are the exact derivatives of the total
energy are straightforwardly obtained in an analytic fashion.
However, in Refs.~5 and 6
convergence with respect to the assumed range of the density
matrix was found to be fairly slow.

This paper presents a generalization of the variational
density-matrix method to general local potentials in
one dimension. In Section II, I describe the mathematical
formalism.  It uses a trial density matrix which is based
on travelling orbitals built out of linear combinations
of harmonic-oscillator eigenfunctions, together with the
``purification scheme'' mentioned above. The method becomes
progressively more accurate as one includes more orbitals
per spatial mesh point. For one basis orbital per mesh point,
one has only one piece of information per mesh point,
and in this sense the theory is mathematically analogous
to the Thomas-Fermi theory (although the kinetic-energy
is a nonlocal functional of the electron density in the present
case, as opposed to a local one in the Thomas-Fermi theory).
As more orbitals are included, one carries more information
per mesh point and thus has a richer description.  In terms
of the coefficients of the travelling basis orbitals, the
total-energy function takes the form of a multicomponent cubic
Landau theory. Density-functional theory has shown that it is
possible to write the total ground-state energy of an electronic
system entirely in terms of the electronic charge density.
However, I do not follow this route here, because obtaining
the kinetic energy in terms of the electron density is very
difficult. In the density-matrix approach, the kinetic energy
is given as a straightforward linear function of the density matrix;
the prices that one pays for this simplicity are that one has
to carry more variables per spatial mesh point, and deal with
constraints that are difficult to implement.

Section III describes applications to model one-dimensional
systems.  These include the noninteracting free-electron gas,
the ``Morse'' potential for bound states, a bimetallic
``interface'' between two different constant potentials,
and a ``semiconductor'' defined by an oscillatory potential.
The applications are intended to illustrate the basic physics
of the method, and to establish whether the new approach contains
several important physical phenomena which are hard to obtain
in real-space total-energy functionals such as Thomas-Fermi
theory or gradient-enhanced versions thereof.  These phenomena
include charge quantization in attractive potentials, Friedel
oscillations from potential perturbations, and band gaps in
semiconductors. I find that the first two are realized in a very
satisfactory fashion.  The band gaps are realized in an approximate
sense, in that a region of reduced electronic density of states
is seen, but no actual band gap.
\newpage
Section IV concludes the paper with an evaluation of the utility
of the method, and a discussion of possible applications to the
inclusion of electron-electron interactions and to embedding clusters
of atoms in media with prescribed boundary conditions.

\section{Mathematical Formulation and Implementation}

The overall procedure is to minimize the energy with respect
to a ``trial'' density matrix $\hat \rho_{\rm tr}$, from which
the variational density matrix $\hat \rho$ entering the total
energy is obtained via a nonlinear ``purification'' transformation.

The underlying variational principle\cite{aec83} states that
the exact zero-temperature density matrix $\hat \rho_{\rm exact}$,
for a system with given chemical potential $\mu$, is the one which
minimizes the functional Tr$(\hat H - \mu \hat I ) \hat \rho$,
subject to the constraints that $\hat \rho$ is real symmetric
and all of its eigenvalues $\lambda$ satisfy $0\leq\lambda\leq 1$.
Note that although in the true density matrix, the eigenvalues are
precisely $0$ and $1$, it is not necessary to specify this as a
constraint for the variational principle; this is instead achieved
automatically by the exact density matrix $\hat \rho_{\rm exact}$
which minimizes the energy functional. In the present case of an
approximate variational density matrix $\hat \rho$, the energy
minimization does not lead to eigenvalues which are precisely
$0$ and $1$, but they are closer to these values than those of
$\hat \rho_{\rm tr}$.  The variational principle, as stated above,
is difficult to use because the eigenvalue constraint is hard
to implement. For this reason, a ``purification'' transformation
has been developed which converts a wide range of trial density
matrices $\hat \rho_{\rm tr}$ into density matrices which are
``allowable'' in the sense that they satisfy the eigenvalue
constraint. The transformation is as follows:
\begin{equation}
\hat\rho =3\hat\rho_{\rm tr}^2 - 2\hat\rho_{\rm tr}^3 ~~~.
\end{equation}
The eigenvalues $\lambda$ of $\hat \rho$ are related to those of
$\hat \rho$, $\lambda_{\rm tr}$, by
\begin{equation}
\lambda =3 \lambda_{\rm tr}^2 -2\lambda_{\rm tr}^3 ~~~,
\end{equation}
so that if all of the $\lambda_{\rm tr}$ are between -1/2 and +3/2,
then all of the $\lambda$ are between $0$ and $1$, and
$\hat \rho$ is allowable. Because $d \lambda /d \lambda_{\rm tr}$
vanishes at  $\lambda_{\rm tr}=0$ and $1$, this transformation has
the tendency to ``pile up'' eigenvalues around $0$ and $1$,
where they belong.

I use the following representation for the trial density
matrix $\hat \rho_{\rm tr}$:
\begin{equation}
   \rho_{\rm tr}(x,x') = \sum_{M=0}^{M_{\rm max}}
    \rho_M (\bar x ) \phi_M (\Delta x)~~~,
\end{equation}
where
\begin{eqnarray}
     \bar x = && (x+x')/2~~~,\\
   \Delta x = && (x-x')~~~,\\
   \phi_M (\Delta x) =
              && (\Delta x/d)^{2M} \exp (-\Delta x^2/2d^2) ~~~,
\end{eqnarray}
and $d$ is a length-scale parameter. The parameter $M_{\rm max}$
determines the number of basis orbitals used in the expansion,
and at fixed $d$ determines the range and number of oscillations
of the density matrix.  Note that the $\phi_M$ are linear combinations
of harmonic-oscillator eigenfunctions.  One can show straightforwardly
from the symmetry $\rho_{\rm tr}(x,x')=\rho_{\rm tr}(x',x)$ that only
even powers are needed in the expansion.

The $\rho_M (\bar x)$ are density functions that, for increasing
$M_{\rm max}$, provide an increasingly accurate description of the
density matrix.
For $M=0$, $\rho_0 (\bar x)$ is simply the charge density corresponding
to $\hat \rho_{\rm tr}$. For $M > 0$, $\rho_M (\bar x)$ determines
the variation of the $\hat \rho_{\rm tr}$ matrix away from the diagonal
points $x = x'$.  The energy functional is obtained from
the $\rho_M(\bar x)$ as follows. The kinetic energy is given by
\begin{equation}
T = (-\hbar^2/2m) \int lim_{x' \rightarrow x}
\bigtriangledown_{x'}^2 \rho(x',x) dx
\end{equation}
which in terms of $\rho_{\rm tr}$ becomes
\begin{eqnarray}
T = && (-\hbar^2/2m)[3\int [\bigtriangledown_x^2\rho_{\rm tr}(x,x')]
       \rho_{\rm tr}(x',x) dx\, dx' \nonumber \\
    && -2\int [\bigtriangledown_x^2 \rho_{\rm tr}(x,x')]
       \rho_{\rm tr}(x',x'') \rho_{\rm tr}(x'',x)dx\, dx'\, dx'']~~~.
\end{eqnarray}
Similarly, one has for the potential energy
\begin{eqnarray}
 U = && \Biggl[ 3 \int V(x) \rho_{\rm tr} (x,x')
        \rho_{\rm tr}(x,x) dx\, dx' \nonumber \\
     && -2 \int V(x) \rho_{\rm tr}(x,x')
        \rho_{\rm tr}(x',x'')\rho_{\rm tr}(x'',x)dx\,dx'\,dx''\Biggr],
\end{eqnarray}
where $V(x)$ is the one-electron potential.  The chemical-potential
contribution to the energy is given by a similar term, but with
$V(x)$ replaced by the constant $\mu$. Since $\rho_{\rm tr}(x,x')$
is linearly related to the $\rho_M (\bar x)$, the total energy $E$
is a cubic functional of the $\rho_M (\bar r)$, and can thus
be written in the form
\begin{eqnarray}
E && = \sum \int E_{MM'}^{(2)} (x,x')
       \rho_M(x) \rho_{M'}(x')dx\, dx' \nonumber \\
  && + \sum \int E_{MM'M''}^{(3)} (x,x',x'')
        \rho_M(x) \rho_{M'}(x') \rho_{M''}(x'')dx\, dx'\, dx''
\end{eqnarray}
where the coefficients $E_{MM'M''}^{(3)} (x,x',x'')$
and $E_{MM'}^{(2)} (x,x')$ are
determined by the basis functions $\phi_M$, $\mu$, and $V(x)$.
This has the form of a multicomponent Landau theory.
The simplicity of this form may be useful in finding
improved algorithms for minimizing the total energy.

The procedure for implementing the formalism is as follows.
One first chooses values of $M_{\rm max}$ and $d$ (appropriate
values are discussed below). One then chooses a mesh for
$\bar x$ and chooses initial values for $\rho_M (\bar x)$
on this mesh. The total number of variables is equal to
$(M_{\rm max}+1)$ times the number of mesh points.  I have typically
used free-electron initial values for $\rho_M (\bar x)$.
The integrals for the energy are evaluated numerically on the
$r$-space mesh.  Because the variational density matrix generated
in this fashion has Gaussian decay at long distances,\cite{densitym}
it is possible to regard it as truncated beyond some critical
radius $R_{\rm max}$.  This means that the numerical integral
for the energy is of order $N$. The energy minimization is
performed using a conjugate-gradient procedure. For this procedure,
one requires the ``generalized forces'', which are the derivatives
of $E$ with respect to the values of $\rho_M(\bar x)$
at mesh points. Because of the simple cubic plus quadratic
form of Eq.~(10), these derivatives are straightforwardly
obtained as numerical integrals similar to those for the energy.
The conjugate-gradient algorithm is allowed to run until the
generalized forces are of order $10^{-4}$ atomic units.

Because the energy  is only a local minimum, not a global one,
some choices of initial conditions and potentials $V(x)$ lead
to ``runaways'' in which the energy becomes negatively infinite
because of unphysical negative occupancies of positive-energy
states. It was found that this problem could be cured by adding
a term proportional to
$\int [\rho_{\rm tr}^3(x,x)-\rho_{\rm tr}^2(x,x)]^2\, dx$.
This term vanishes if the eigenvalues of $\hat \rho_{\rm tr}$
are precisely 0 or 1.  The actual energy corresponding
to this term is quite small, since the eigenvalues of the
minimizing $\hat \rho_{\rm tr}$ are all close to 0 or 1,
but the term appears to prevent the runaways consistently.

\section{Applications}

The main purpose of the applications is to establish the extent
to which the new method obtains inherently quantum-mechanical
effects that are not readily obtained in real-space total-energy
methods, such as Thomas-Fermi type theories. These effects are
charge quantization at localized potentials, Friedel oscillations
around scatterers, and band gaps in semiconductors. At localized
potentials, the Thomas-Fermi theory obtains a total charge that
varies continuosly with the chemical potential; the correct
quantum-mechanical charge varies discontinuously with the chemical
potential, with the discontinuities occurring at bound-state energies.
Around scattering potentials, the Thomas-Fermi theory obtains
a smooth exponentially decaying charge density; the correct density
has a power-law decay with an oscillating prefactor. The density of
states in Thomax-Fermi theory, defined as the derivative of the number
of electrons with respect to the chemical potential, never displays
gaps in periodic potentials, but rather is closely related to the
free-electron density of states.

In all of the applications, I use $M_{\rm max} = 8$ and $d=3a_0$,
where $a_0$ is the Bohr radius. This value of $d$, on the basis
of trial calculations, provides a good compromise between a correct
description at short distances and the need to obtain a sufficiently
long range for the density matrix. The choice $R_{\rm max}=7d$
was then found to lead to numerically converged integrals.
The value of $M_{\rm max}$ was the highest value that I was able
to use without running into numerical difficulties involving
cancellations between large terms in Eq.~(3).

To obtain the simplest picture of the  physical significance of
the approximations of the method, I begin with the one-dimensional
free-electron gas, with the chemical potential $\mu =1$~Ry.
This corresponds to $k_F=1a_0^{-1}$. The density matrix $\hat\rho$
(which depends only on $(x-x')$) is shown in Fig.~1, along
with $\hat\rho_{\rm tr}$ and the exact density matrix
$\rho_{\rm exact}(x,x')=\sin {[k_F(x-x')]}/\pi (x-x')$.
At small distances, both $\hat\rho$ and $\hat\rho_{\rm tr}$ are in
excellent agreement with $\hat\rho_{\rm exact}$.  The
good agreement persists out until about 10~$a_0$. Beyond
this point, $\hat\rho$ becomes increasingly damped with respect
to the $\hat\rho_{\rm exact}$, although substantial oscillations
are still seen out to $20 a_0$ and beyond. Note that
$\hat\rho_{\rm tr}$ decays more rapidly then $\hat\rho$,
so that even on the expanded scale of Fig.~1b, the oscillations
are almost invisible beyond 17~$a_0$. This difference between
$\hat\rho$ and $\hat\rho_{\rm tr}$ is due to the purification
procedure; the convolution implicit in Eq.~(1) serves to increase
the range of $\hat\rho$ beyond that of $\hat\rho_{\rm tr}$. However,
the purification makes up for only about half of the difference
between $\hat \rho_{\rm tr}$ and $\hat \rho_{\rm exact}$.

Figure 2 shows the kinetic and total energy densities for
the electron gas, as functions of $M_{\rm max}$.  At
$M_{\rm max}=0$ (only one variable per mesh point),
the results are quite inaccurate in comparison with the
exact values. However, already for $M_{\rm max} = 2$,
considerable improvements are seen, and for $M_{\rm max}=8$,
the agreement is quantitatively accurate.

I now turn to the case of a bound state. Because of its analytic
tractability, I choose the potential
\begin{equation}
V(x) = {-\hbar^2 \kappa^2 \over 2m}\,
       {2 \over \cosh^2 \kappa r}~~~.
\end{equation}
This potential has one bound state wave function,
$\psi (x) = \sqrt {\kappa /2} (1/\cosh {\kappa x})$,
with eigenvalue $-\hbar^2 \kappa^2/2m$. I consider the
case $\kappa = 0.5 a_0^{-1}$ and use $\mu = -0.1$~Ry,
in comparison with the bound state energy of $-0.25$~Ry.
The charge density for $M_{\rm max}=8$ is shown in Fig.~3.
On the scale of the figure, it is indistinguishable from the
exact one. For this value of $\mu$, the total charge
$Q = \int_{-\infty}^{+\infty} \rho (x,x) dx$ is
1.0001, which is thus in error by only one part in $10^{4}$.
The bound-state energy of $-0.2495$~Ry is also very close
to the exact value.

To explore in more detail the nature of the charge quantization,
Fig.~4 shows $Q$ as a function of $\mu$. The exact $Q$
jumps from $0$ to $1$ at $-0.25$~Ry. The $Q$ obtained from
the density matrix follows this behavior closely, except
that it climbs to 1 over a narrow but finite range from about
$-0.26$~Ry to $-0.23$~Ry. Above $-0.23$~Ry, $Q$ is very close
to constant.

The entire density matrix for $\mu = -0.1$~Ry is compared to
$\rho_{\rm exact}(x,x')=\psi (x)\psi (x')$ in Fig.~5.
The contour plots are essentially indistinguishable, the only
visible difference being that the contours for the variational
density matrix are somewhat more square close to the origin.
Thus all aspects of this bound-state problem seem to be described
quite well by the approximate variational density matrix.

Our model for the one-dimensional model metallic interface
has the step-function form
\begin{eqnarray}
  V(x) = V_- \qquad (x<0) ~~~, \nonumber\\
{\rm and}\qquad\qquad\qquad
  V(x) = V_+ \qquad (x>0)~~~.
\end{eqnarray}
I choose $V_-=-1$~Ry and $V_+= -2$~Ry, and consider
the case $\mu = -0.5$. The corresponding charge densities
are indicated in Fig.~6.  The approach to the bulk densities
on either side of the interface is oscillatory as expected
according to the standard Friedel-oscillation theory.
The wavelengths obtained by the variational density matrix
are close to the expected wavelengths $(\pi /k_F)$ which have
the values $4.4$\AA\, on the left and $2.6$\AA\, on the right.
The main point of difference between the variational density
matrix results and the exact ones is that the oscillations
in the former case eventually have a Gaussian decay,
while the exact calculation gives a decay proportional
to $\sin 2k_Fx/x$.

Our last example is a model semiconductor, defined by the potential
\begin{equation}
  V(x) = 2V_0 \cos {qx} ~~~.
\end{equation}
As is well known, in a weak-scattering analysis this type
of potential produces a band gap of magnitude $|2V_0|$,
centered around the kinetic energy $E_0 = \hbar^2 (q/2)^2/2m$.
Around the band gap, the density of states $g(E)$ per unit
length (in this one-dimensional case) displays singularities
of the form $g(E) \propto 1/\sqrt{(|(E-E_{c,v})|}$,
where $E_{c,v}$ indicates the conduction- and valence-band edges.
Although the density matrix does not give the density of states
directly, we can obtain $g(E)$ by evaluating the dependence
of the total charge $Q_{\rm tot}$ on the chemical potential:
\begin{equation}
  g(E)L = dQ_{\rm tot}/d\mu ~~~,
\end{equation}
where $L$ is the system size. We use the parameters $V_0=0.15$~Ry
and $q=2a_0^{-1}$, which in the weak-scattering theory would lead
to a gap of width $0.3$~Ry centered about $1$~Ry. The calculated
density of states is shown in Fig.~7, and is compared with the
exact free-electron density of states. It is seen that in a region
extending from about $0.87$~Ry to $1.20$~Ry, the density of states
is considerably reduced. At $0.95$~Ry, it is roughly six times
smaller than the free-electron value. I term this effect
a ``quasigap''. The width of the quasigap is comparable to the
free-electron prediction. Also, on either side of the gap,
pronounced increases in $g(E)$ are seen, which are presumably
connected with the square-root singularities of the true density
of states. I do not know whether this variational density-matrix
method actually can obtain square-root singularities.

\section{Conclusion}

The main conclusion to be drawn from the above is that there
exists an order-$N$ variational density-matrix method for
calculating electronic structure, which
obtains quantitative agreement with exact results in several
cases. It contains several effects that are hard to extract
from real-space density-based descriptions: charge quantization,
Friedel oscillations, and band-gap formation.  One could
straightforwardly combine such a method with the local-density
approximation (LDA) of density-functional theory by simply adding
additional terms to the Hamiltonian, to get a viable total-energy
method.  However, the use of the density matrix rather than the
density as a basic variable may also make it feasible to develop
improvements on the LDA. It is probably easier develop a picture
of the electronic pair correlations in the system from the density
matrix rather than the density itself. For example, one knows that
in insulators and semiconductors, the density matrix decays
exponentially, with a decay rate determined by the band gap.
Thus knowledge of the decay rate of the density matrix may give
some information about the excited-state spectrum of a material.

I believe that another advantage of a method such as this one
is that, because it uses an $r$-space representation, one can
easily embed a calculation for a strongly distorted or disrupted
piece of material into a host of material which is essentially
perfect. One can simply specify that the $\rho_M(\bar x)$ in
the perfect region have their perfect-lattice behavior, and
allow them to vary arbitrarily in the disrupted region, subject
to the boundary conditions. In this way, one can avoid
the use of periodic boundary conditions, which are typically
necessary in $k$-space representations.

The main hurdle to be treated before the extension to three
dimensions and the inclusion of electron-interaction terms is
to streamline the procedure to obtain greater computational
efficiency. I find that with standard conjugate-gradient
methods, achieving convergence with 900 variables (100
mesh points, nine variables per mesh point) takes several
minutes of computer time on a Silicon Graphics R4000
workstation. At this speed, doing any but the simplest
three-dimensional problems would be computationally prohibitive.
Two avenues are likely to help. The first is to speed up
the numerical integrals. The computer time is dominated
by the numerical integrals that are done in order to compute
$\hat \rho_{\rm tr}^3$. It is possible that these can be speeded
up by using an intermediate expansion step, in which
$\hat \rho_{\rm tr}^2$ is expanded in a form analogous to Eq.~(3)
for $\hat \rho_{\rm tr}$. If such an expansion can be made, then
the time required to compute $\hat \rho_{\rm tr}^3$ could be reduced
by an order of magnitude. The second avenue that might help
is a speeding-up of the line searches in the conjugate gradient
procedure. Since the total energy is a cubic function of
the underlying variables, the minimum along the line search
can be precisely determined by knowing the forces at three
points along the line.\cite{vanderbilt}

\acknowledgments

I am grateful to Murray Daw and David Vanderbilt for useful
comments and advice, and to Lilly Canel for a careful reading
of the manuscript.  This work was supported by the
Department of Energy under Grant Number DE-FG02-84ER45130.

\begin{figure}
\caption{Comparison of density matrices for one-dimensional free-electron
        gas. $\rho_{\rm tr}$ denotes ``trial'' density matrix that determines
        variational density matrix via the purification transformation.}
\label{fig1}
\end{figure}
\begin{figure}
\caption{Estimates of kinetic-energy and total-energy densities for
        one-dimensional free-electron gas, obtained by variational
        density matrix. $2M_{\rm max}+1$ is the number basis orbitals
        per mesh point used in expanding the density matrix.}
\label{fig2}
\end{figure}
\begin{figure}
\caption{Charge density $\rho(x,x)$ and one-electron potential
        $U(x)$ for Morse potential well, using variational density matrix.}
\label{fig3}
\end{figure}
\begin{figure}
\caption{Total charge vs. chemical potential for Morse bound-state
        potential, using variational density matrix.}
\label{fig4}
\end{figure}
\begin{figure}
\caption{Contour plot of density matrix for Morse bound-state potential,
        using variational density matrix (a), in comparison with exact
        one (b).}
\label{fig5}
\end{figure}
\begin{figure}
\caption{Charge density for step-function model of bimetallic interface,
        obtained using variational density-matrix method.}
\label{fig6}
\end{figure}
\begin{figure}
\caption{Density of states for model semiconductor (potential defined
        in text), using variational density-matrix method. ``Free-Electron''
        denotes density of states in absence of scattering potential. }
\label{fig7}
\end{figure}
\newpage
\epsfbox{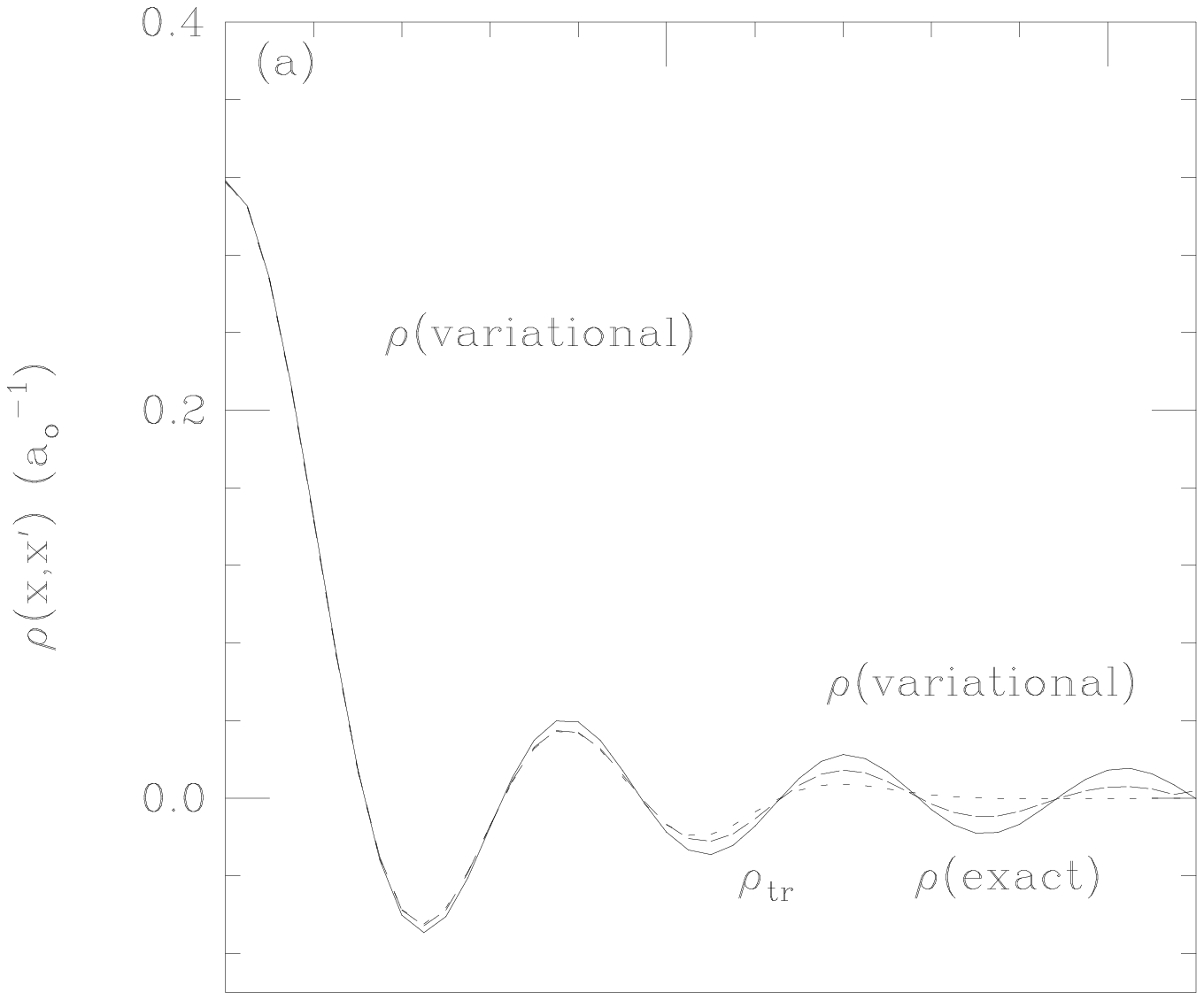}
\epsfbox{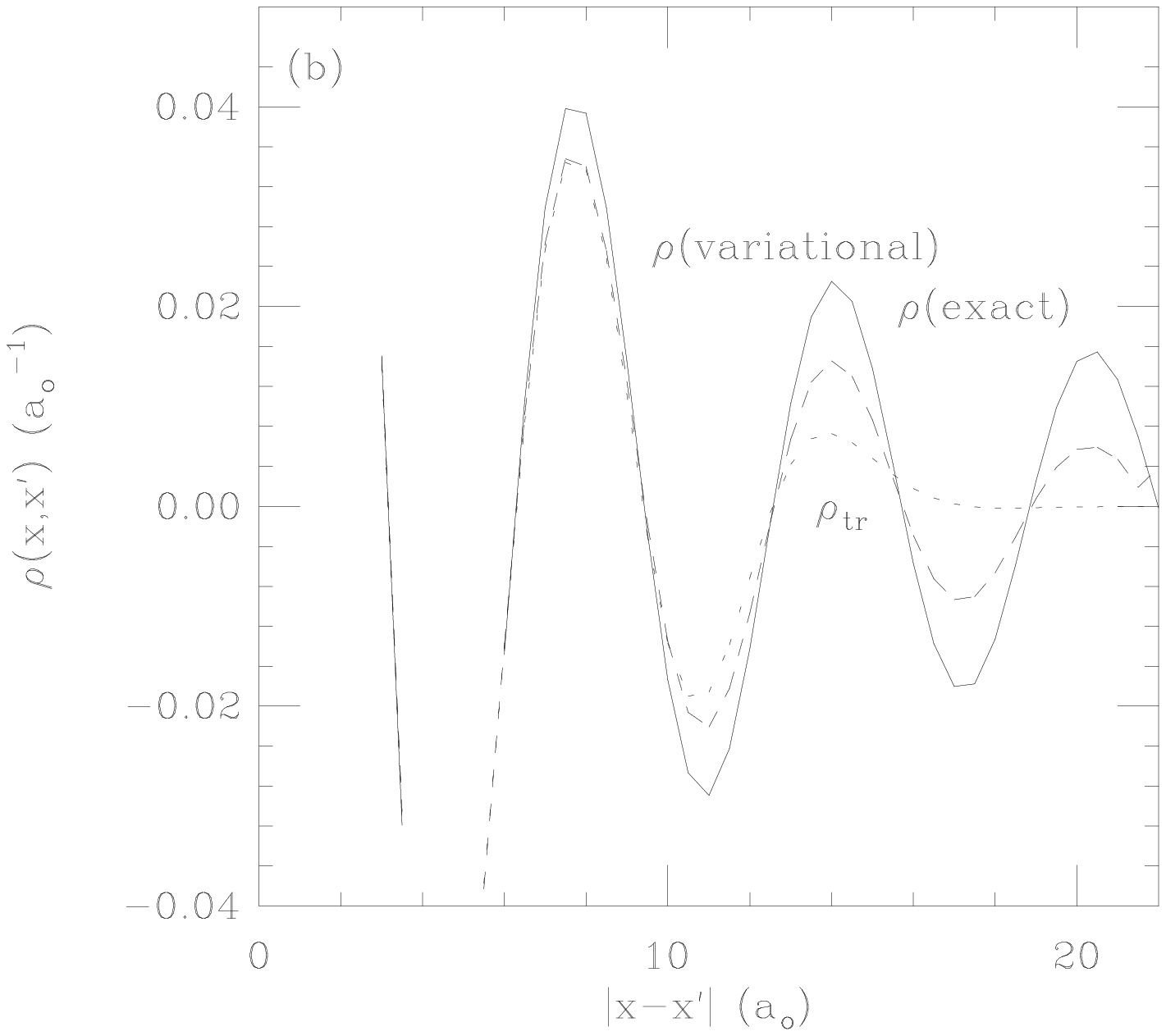}
\newpage
\epsfbox{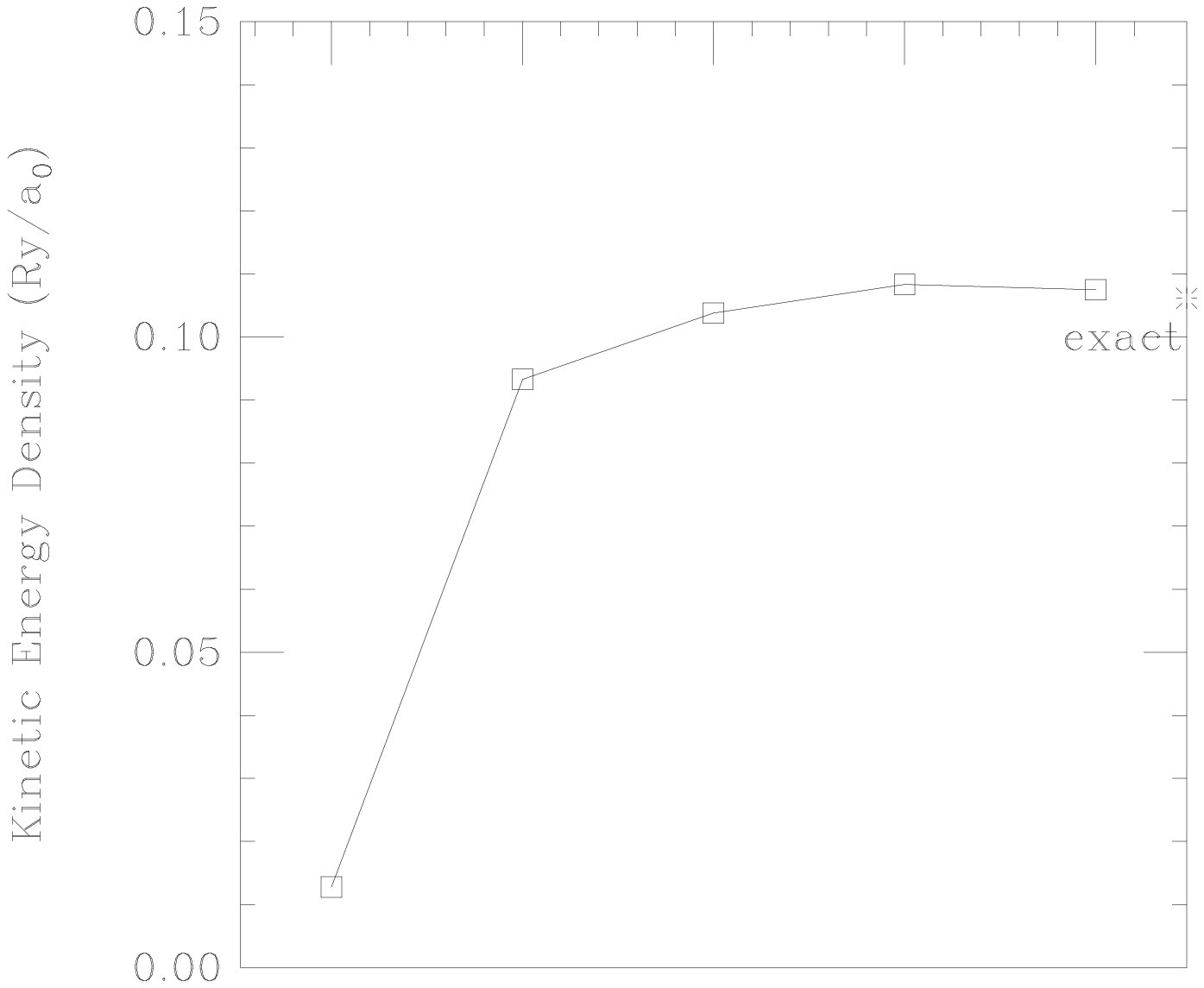}
\epsfbox{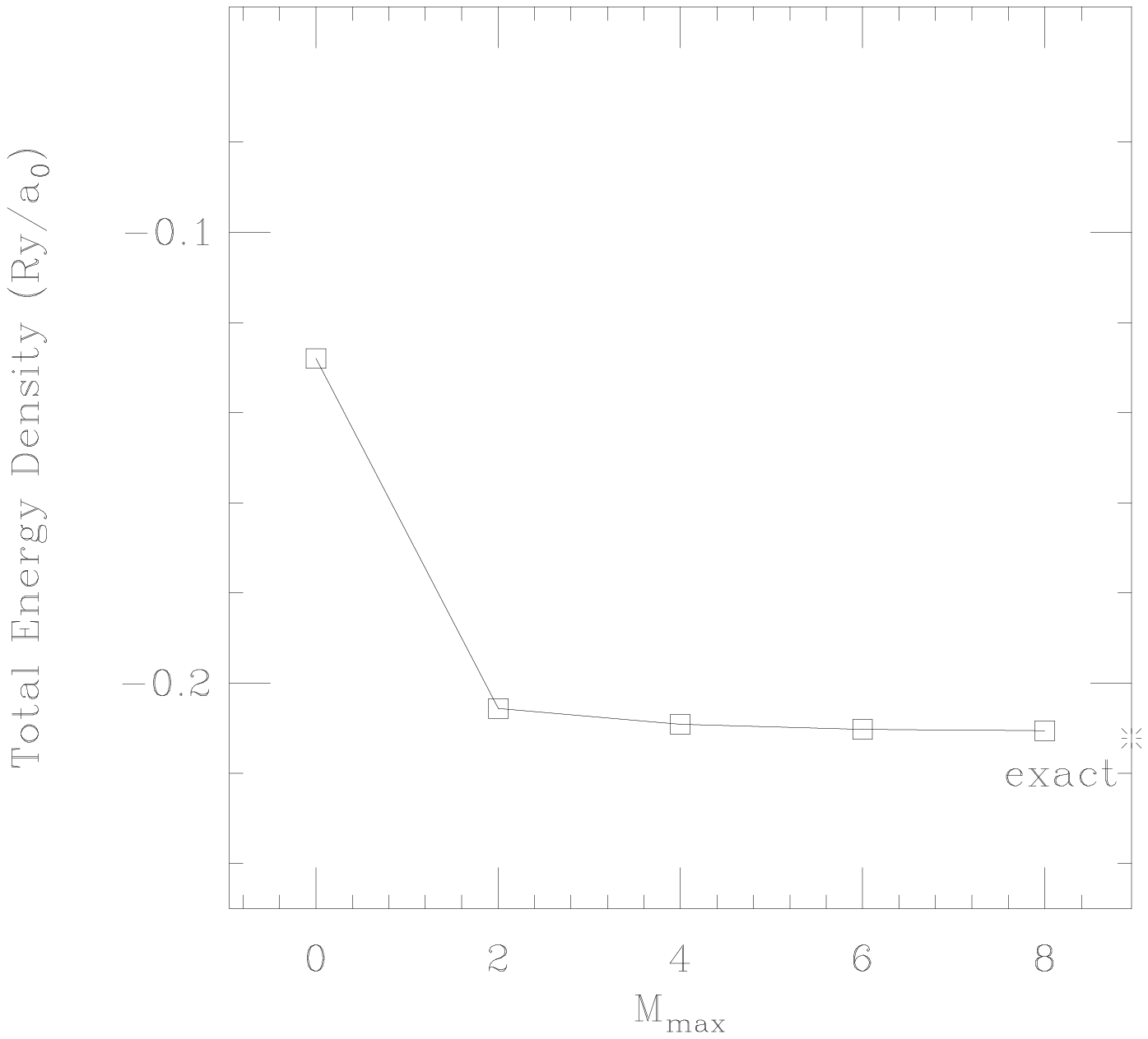}
\newpage
\epsfbox{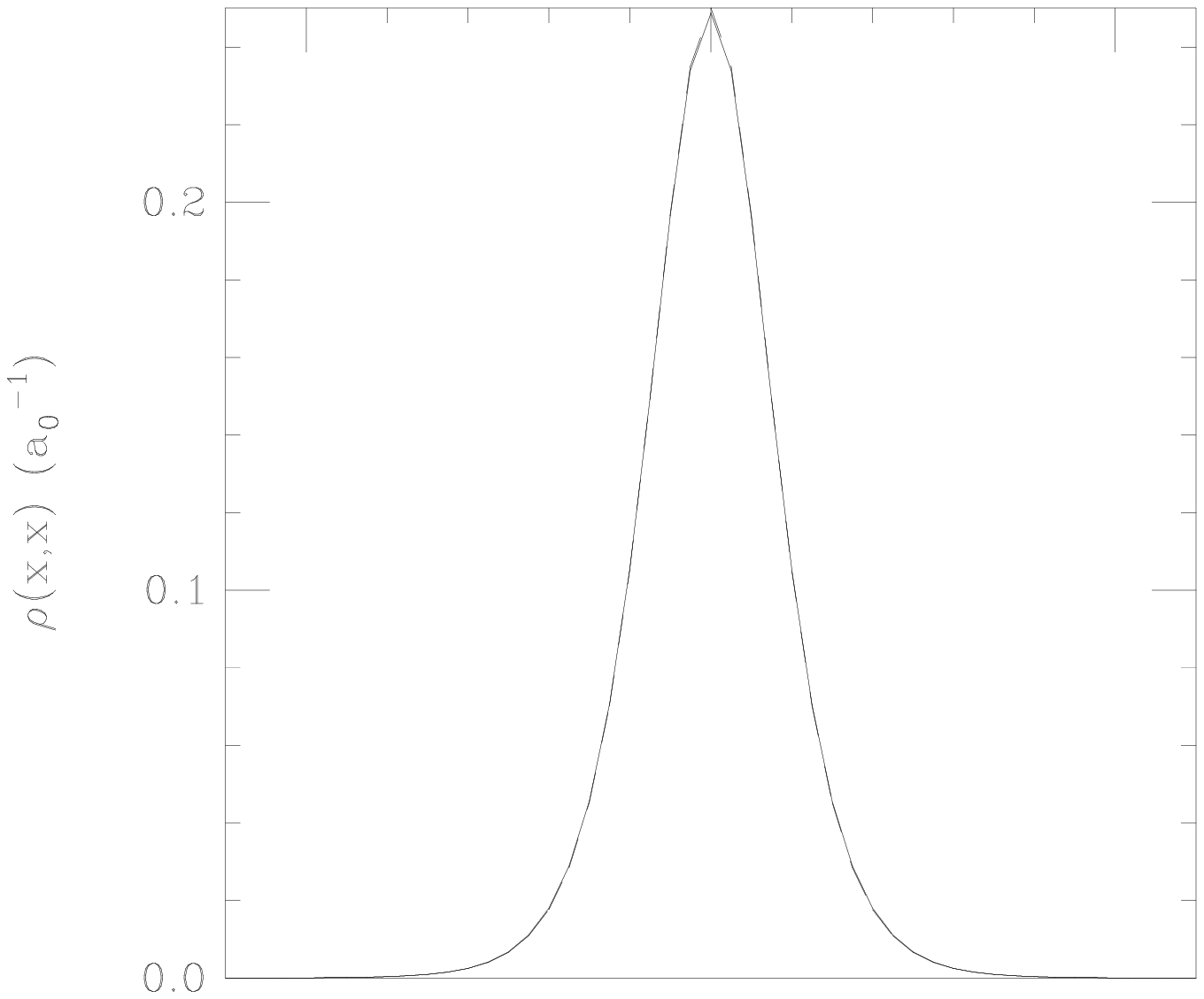}
\epsfbox{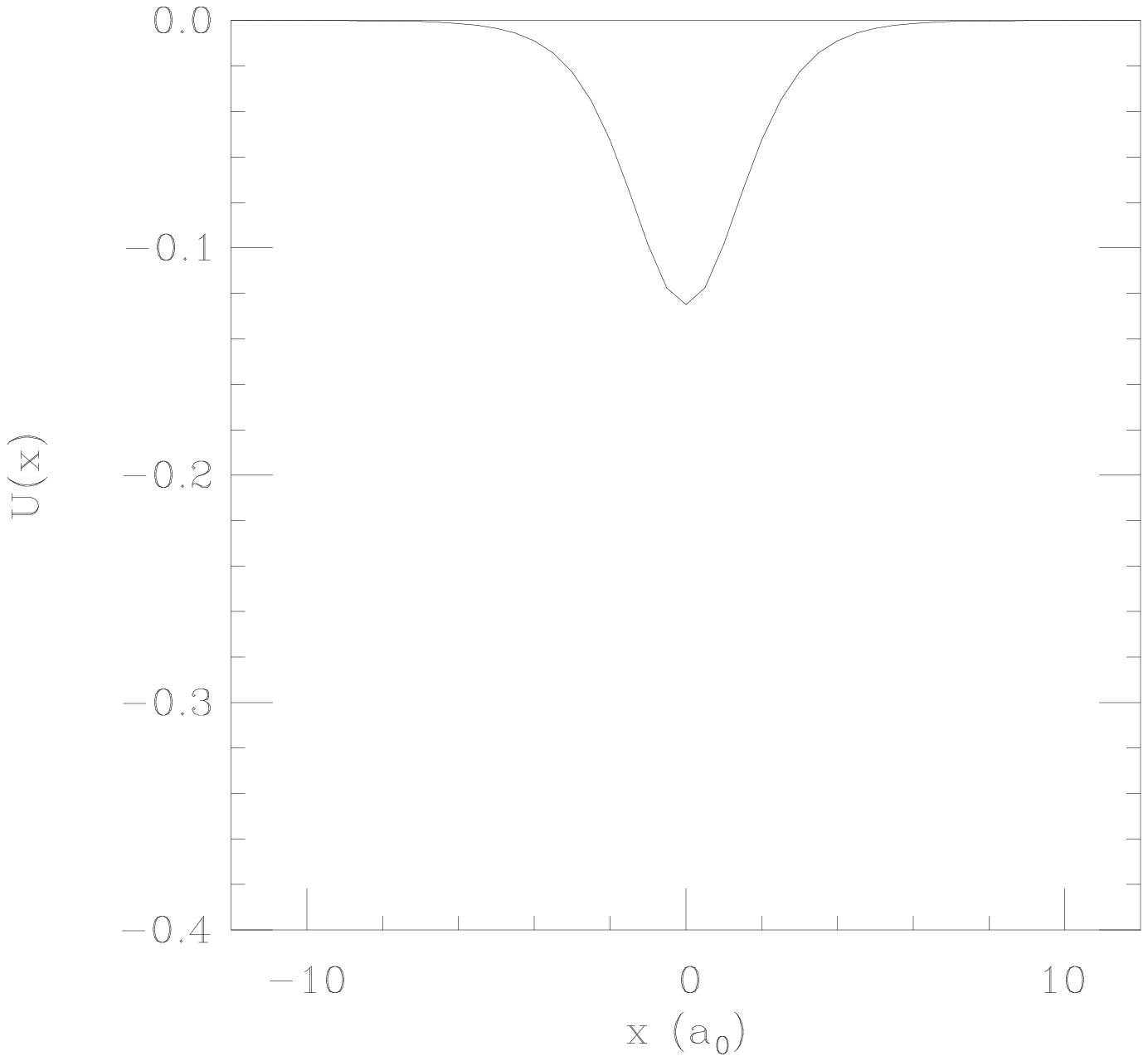}
\newpage
\epsfbox{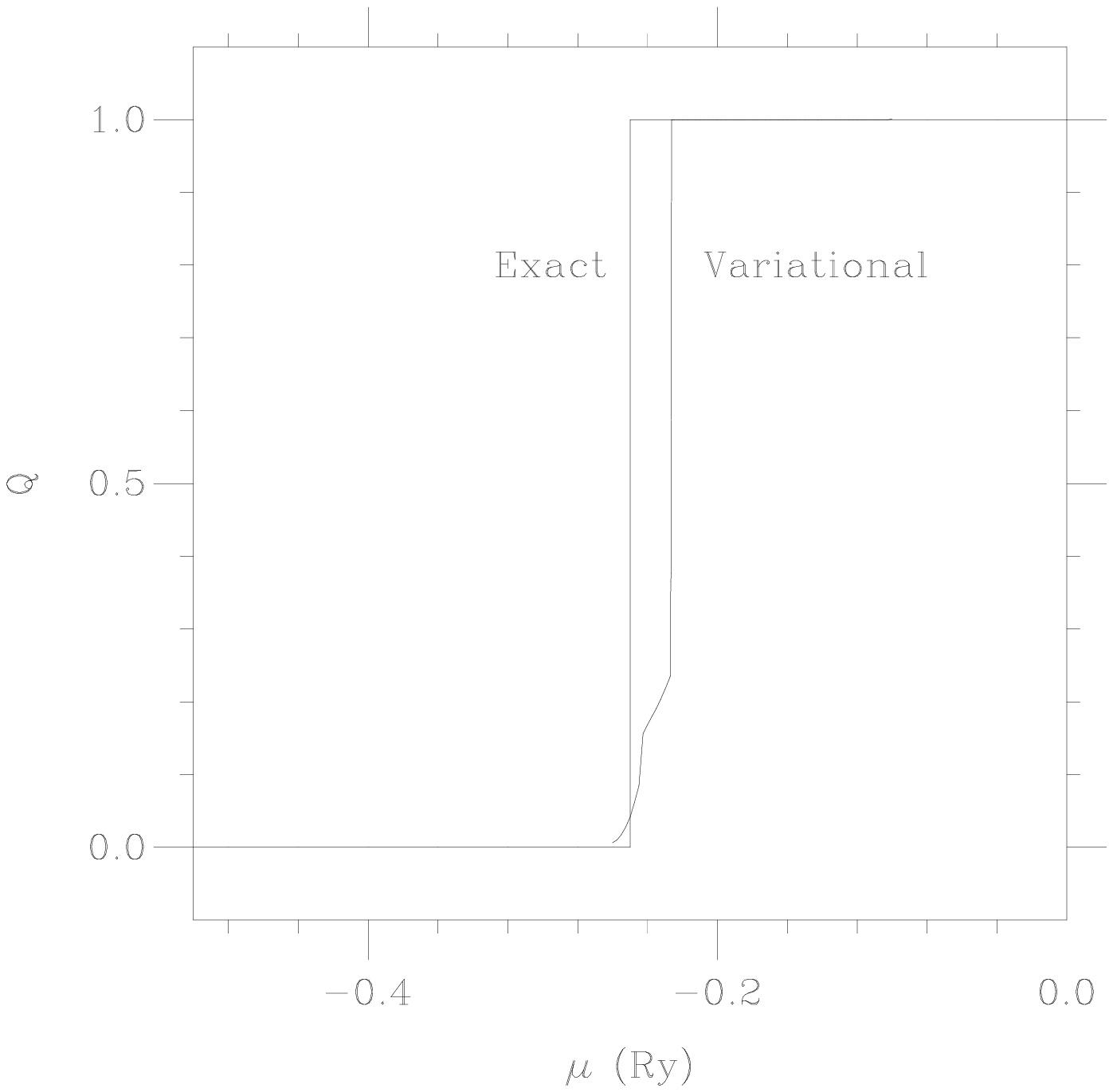}
\newpage
\epsfbox{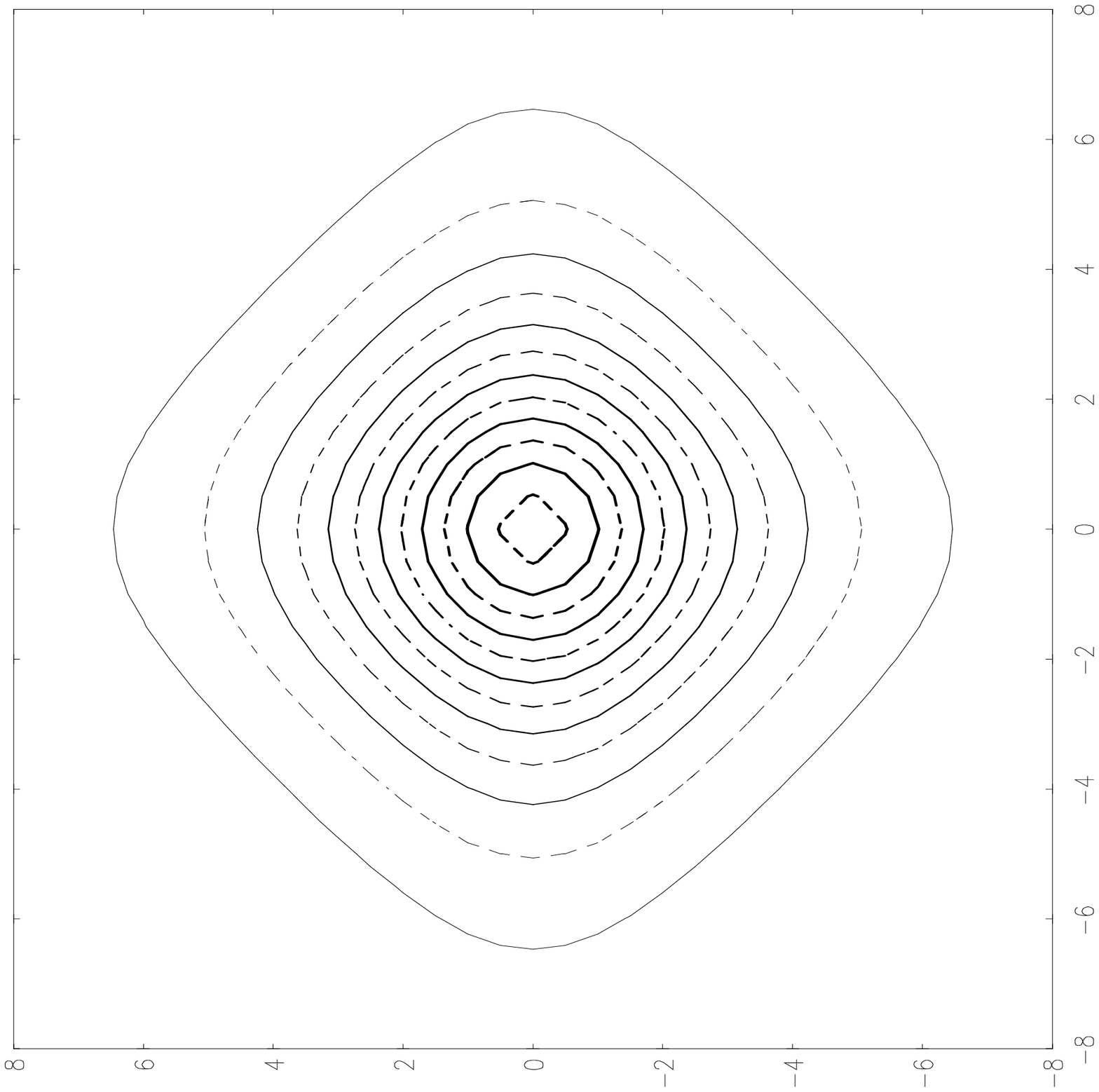}
\epsfbox{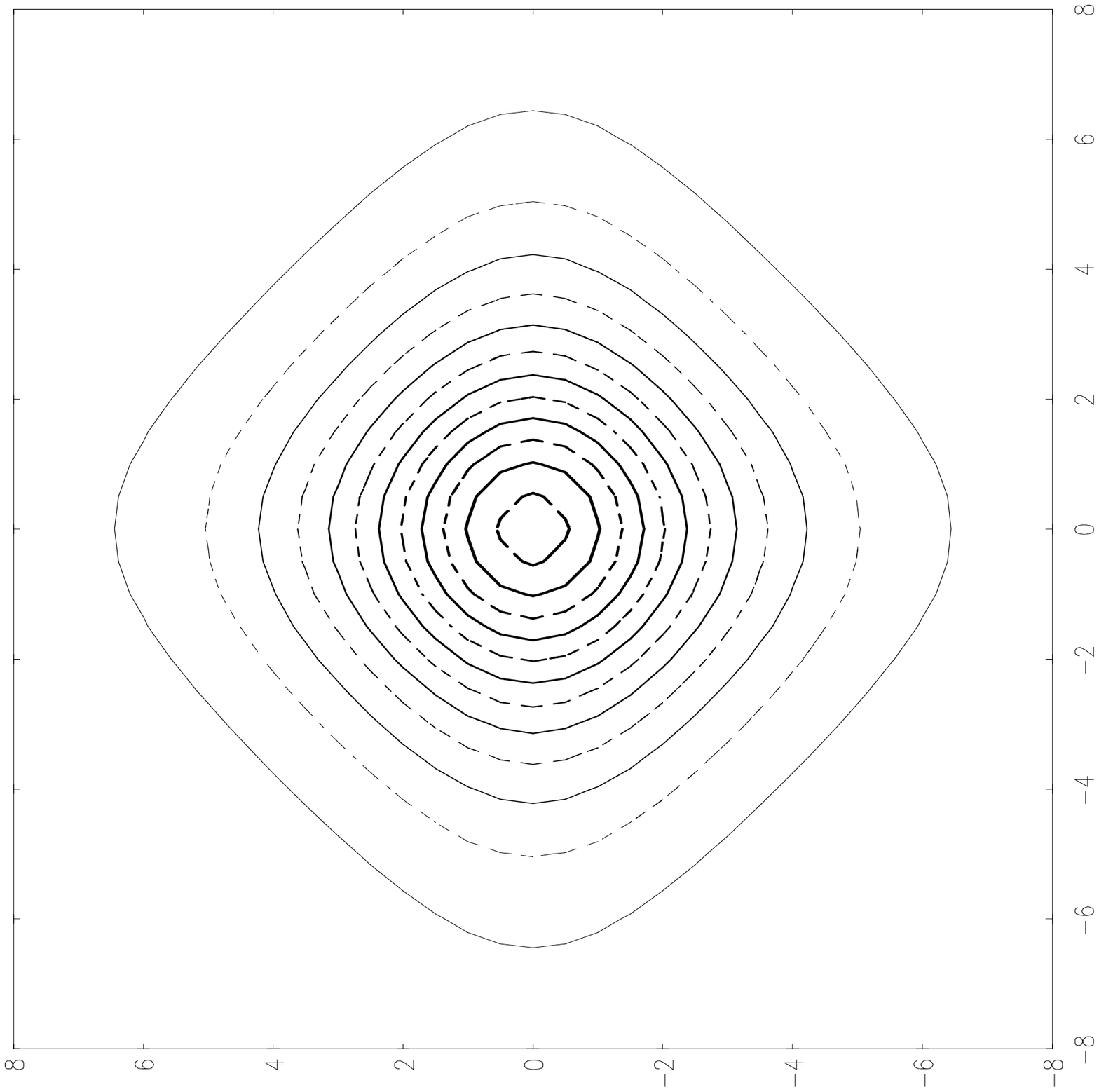}
\newpage
\epsfbox{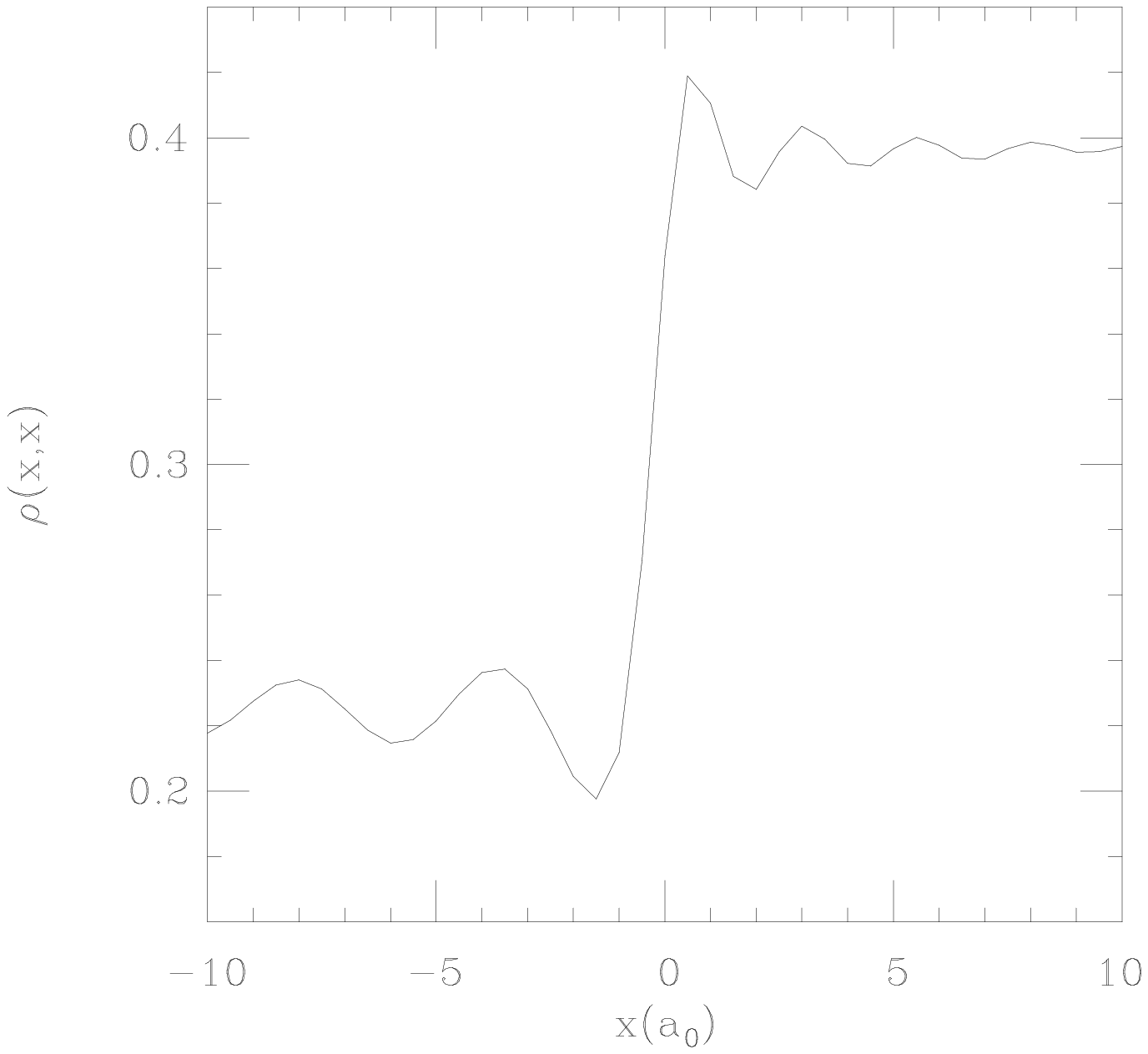}
\newpage
\epsfbox{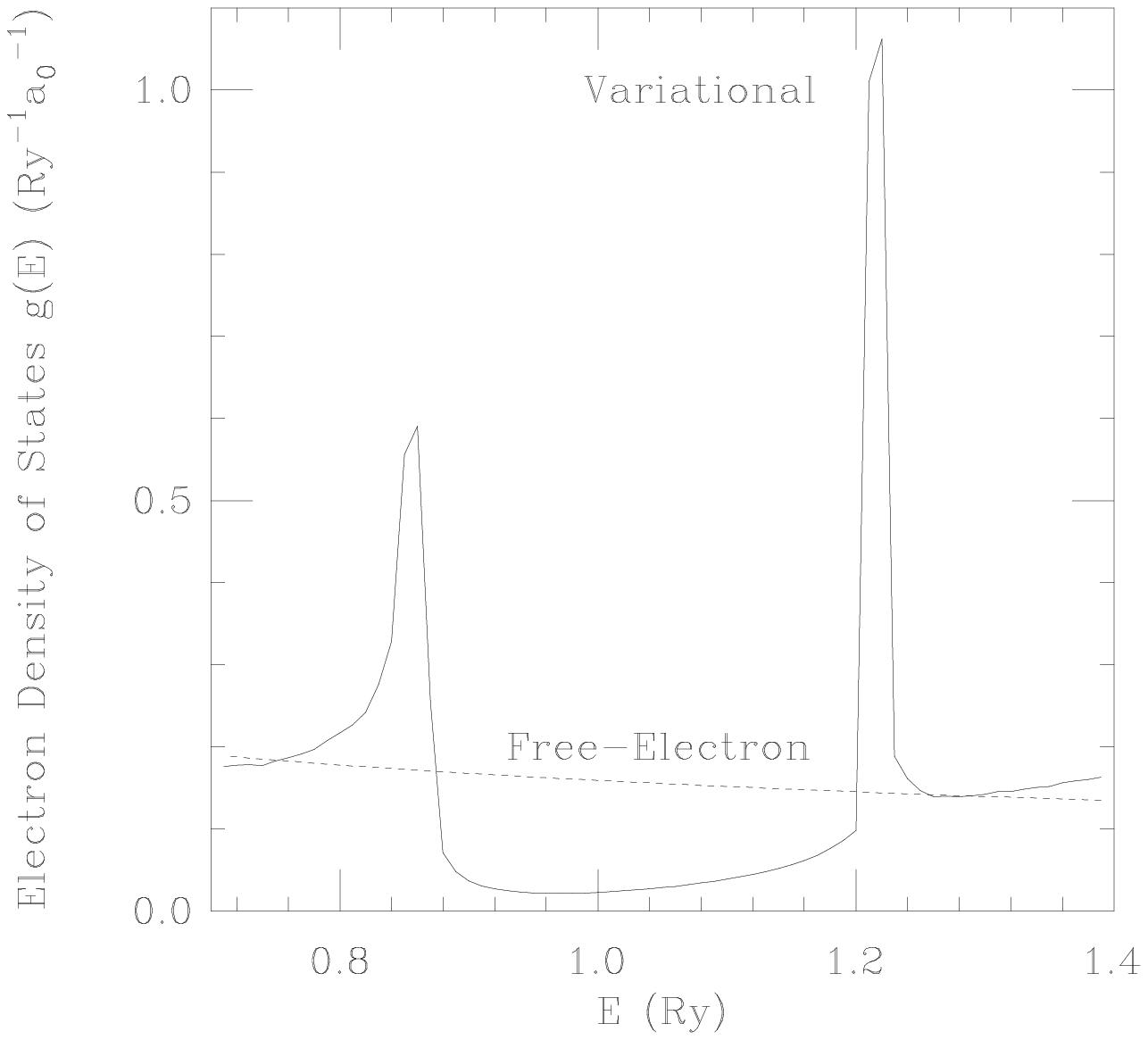}

\begin{references}
\bibitem{wyang91}
W. Yang, Phys. Rev. Lett. {\bf 66}, 1438 (1991).
\bibitem{baroni92}
S. Baroni and P. Giannozzi, Europhys. Lett. {\bf 17}, 547 (1992).
\bibitem{galli92}
G. Galli and M. Parrinello, Phys. Rev. Lett. {\bf 69}, 3547 (1992).
\bibitem{stechel94}
E. B Stechel, A. R. Williams, and P. J. Feibelman,
Phys. Rev. {\bf B 49}, 10088 (1994).
\bibitem{xpLi93}
X.-P. Li, R. W. Nunes, and D. Vanderbilt, Phys. Rev. {\bf B 47},
10891 (1993).
\bibitem{daw93}
M. S. Daw, Phys. Rev. {\bf B 47}, 10899 (1993).
\bibitem{mcweeny60}
R. McWeeny, Rev. Mod. Phys. {\bf 32}, 335 (1960).
\bibitem{smith75}
J. R. Smith and J. G. Gay, Phys. Rev. {\bf B 12}, 4238 (1975).
\bibitem{aec83}
A. E. Carlsson and N. W. Ashcroft, Phys. Rev. {\bf B 27}, 2101 (1983).
\bibitem{densitym}
This implies that the asymptotic behavior of the density matrix,
for any finite $M_{\rm max}$, is wrong. However, for sufficiently
large $M_{\rm max}$, the short-ranged part of the density matrix
is obtained very well.
\bibitem{vanderbilt}
David Vanderbilt (private communication).

\end{references}
\end{document}